\documentclass[fleqn,usenatbib]{mnras}

\usepackage{newtxtext,newtxmath}
\usepackage{gensymb}

\usepackage[T1]{fontenc}

\DeclareRobustCommand{\VAN}[3]{#2}
\let\VANthebibliography\thebibliography
\def\thebibliography{\DeclareRobustCommand{\VAN}[3]{##3}\VANthebibliography}

\usepackage{graphicx}	
\usepackage{amsmath}	
\usepackage{cleveref}	
\crefname{figure}{Fig.}{Figs.}
\crefname{equation}{equation}{equations}

\usepackage{adjustbox}

\newcommand{\um}{$\mu\mathrm{m}$}
\newcommand{\id}{GNWY-7379420231}
\newcommand{\cii}{[C\textsc{ii}]}
\newcommand{\sfrcii}{$\mathrm{SFR}_\mathrm{[C\textsc{ii}]}$}

\newcommand{\orcidsymb}[2]{#1\href{http://orcid.org/#2}{\adjustbox{trim={-.15\width} {0\height} {-.15\width} {0\height},clip}{\includegraphics[height=10pt]{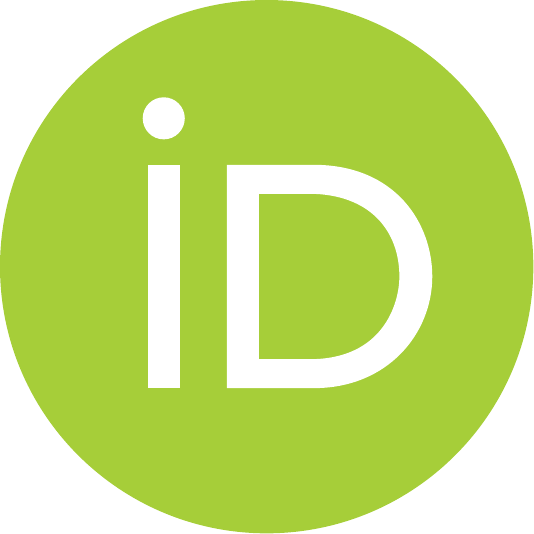}}}}

\title[{[C\textsc{ii}]} emission in a $z=7.1$ UV Bump Galaxy]{NOEMA probes the [C\textsc{ii}] and dust content in a 2175{\AA} UV Bump Galaxy at $z=7.1$}

\author[K. Ormerod et al.]{\orcidsymb{Katherine Ormerod}{0000-0003-2000-3420}$^{1}$\thanks{E-mail: \href{mailto:arikorme@ljmu.ac.uk}{arikorme@ljmu.ac.uk}},
\orcidsymb{Renske Smit}{0000-0001-8034-7802}$^{1}$,
\orcidsymb{Joris Witstok}{0000-0002-7595-121X}$^{2,3}$,
\orcidsymb{Anna de Graaff}{0000-0002-2380-9801}$^{4, 5}$\thanks{Clay Fellow}, \orcidsymb{Michael V.\ Maseda}{0000-0003-0695-4414}${^6}$, \newauthor
\orcidsymb{Irene Shivaei}{0000-0003-4702-7561}${^7}$,
\orcidsymb{Andrew J.\ Bunker }{0000-0002-8651-9879}${^8}$,
\orcidsymb{Gareth C.\ Jones}{0000-0002-0267-9024}${^{9,10}}$
\\
$^{1}$Astrophysics Research Institute, Liverpool John Moores University, 146 Brownlow Hill, Liverpool, L3 5RF, UK \\
$^{2}$Cosmic Dawn Center (DAWN), Copenhagen, Denmark \\
${^3}$Niels Bohr Institute, University of Copenhagen, Jagtvej 128, DK-2200, Copenhagen, Denmark\\
${^4}$Max-Planck-Institut f\"ur Astronomie, K\"onigstuhl 17, D-69117 Heidelberg, Germany \\
$^5$Center for Astrophysics $|$ Harvard \& Smithsonian, 60 Garden St., Cambridge MA 02138 USA \\
$^6$Department of Astronomy, University of Wisconsin-Madison, 475 N. Charter St., Madison, WI 53706, USA\\
${^7}$Centro de Astrobiología (CAB), CSIC-INTA, Ctra. de Ajalvir km 4, Torrejón de Ardoz, E-28850, Madrid, Spain \\
${^8}$Department of Physics, University of Oxford, Denys Wilkinson Building, Keble Road, Oxford OX1 3RH, UK\\
${^9}$Kavli Institute for Cosmology, University of Cambridge, Madingley Road, Cambridge, CB3 0HA, UK\\
${^{10}}$Cavendish Laboratory, University of Cambridge, 19 JJ Thomson Avenue, Cambridge, CB3 0HE, UK
}

\date{Accepted XXX. Received YYY; in original form ZZZ}

\pubyear{\the\year{}}

\begin{document}
\label{firstpage}
\pagerange{\pageref{firstpage}--\pageref{lastpage}}
\maketitle

\begin{abstract}    
The detection of the $2175${\AA} UV bump at $z>6$ challenges existing models of dust formation, suggesting rapid formation of small carbonaceous dust grains within the first billion years of cosmic time. We present the results of the first direct attempt at linking far-infrared (FIR) observations to the UV bump within the Epoch of Reionisation (EoR), through  Northern Extended Millimeter Array (NOEMA) observations of GNWY-7379420231 at $z=7.108$, a galaxy exhibiting the strongest known UV bump feature at $z>4$. 
We detect the [C$\textsc{ii}$] 158$\mu$m emission line at 5.1$\sigma$ at  $z_\mathrm{[C\textsc{ii}]} = 7.1078 \pm 0.0005$, in excellent agreement with the redshift derived from the [O$\textsc{iii}$] $\lambda 5007${\AA} line. 
The {\cii} luminosity implies  $\mathrm{SFR}_\mathrm{[C\textsc{ii}]}=16.2^{+5.8}_{-5.5}M_\odot \mathrm{yr}^{-1}$, consistent with short timescale SFR tracers such as dust corrected $\mathrm{SFR}_\mathrm{H\alpha}=18.0\pm3.9 M_\odot \mathrm{yr}^{-1}$ and $\mathrm{SFR}_\mathrm{10~Myr}=20.5^{+3.8}_{-5.0}M_\odot \mathrm{yr}^{-1}$ from SED fitting. 
The dust continuum is not detected suggesting an obscured SFR of $\mathrm{SFR}_\mathrm{IR} < 32 ~M_\odot \mathrm{yr}^{-1}$ and a dust mass of $M_\mathrm{d} < 5.3\times10^6 ~M_\odot$ $(M_\mathrm{d}/M_\star<2\%)$. Finally, the [C$\textsc{ii}$]-derived dynamical mass of $\log_{10}(M_\mathrm{dyn}/M_\odot)=8.95^{+0.51}_{-0.66}$ and stellar mass of $\log _{10}\left(\mathrm{M}_{\star} / \mathrm{M}_{\odot}\right) = 8.39_{-0.09}^{+0.13}$  suggest a gas-rich moderately massive galaxy. Taken together these results rule out GNWY-7379420231 being a heavily dust obscured or massive galaxy, but rather a `normal' EoR galaxy  with a recent upturn in star-formation. Our results suggest efficient shattering of larger dust grains in the diffuse, turbulent interstellar medium (ISM) and/or fortunate line of sight alignment are needed to explain the UV bump properties of GNWY-7379420231.  

\end{abstract}




\section{Introduction}
Although dust comprises $\leq1\%$ of the mass of a galaxy's interstellar medium (ISM), it is a fundamental component that shapes both the physical and observational properties of galaxies. Dust absorbs approximately half of the optical and ultraviolet (UV) light and re-emits the absorbed energy as infrared (IR) light, with important implications for the observational properties of galaxies \citep[e.g.,][]{kennicutt_star_2012,galliano_interstellar_2018, schneider_formation_2024, Smit_Bowler_2026}. Moreover, dust drives key processes in galaxy evolution by catalysing the formation of molecules and fragmentation of gas clouds, processes which are vital to star formation \citep{schneider_fragmentation_2006, chen_populating_2018}.

The properties of dust in galaxies can be probed through extinction curves, measured along individual sightlines, or attenuation curves, which describe the integrated light within an aperture. Attenuation curves also incorporate additional effects which arise from star-dust geometry within galaxies, such as scattering back into the line of sight (LOS) and the contribution from unobscured stars \citep{narayanan_theory_2018, salim_dust_2020}. Commonly used examples include the Calzetti attenuation curve \citep{calzetti_dust_1994, calzetti_dust_2000} derived for local starburst galaxies, and the Milky Way \citep[MW;][]{cardelli_relationship_1989}, Small Magellanic Cloud (SMC), and the Large Magellanic Cloud (LMC) extinction curves \citep{fitzpatrick_analysis_1986, gordon_quantitative_2003, gordon_expanded_2024}. 
Notably, the MW extinction curve exhibits a prominent `UV bump' feature at $2175${\AA} which was first identified in MW sightlines by \citet{stecher_interstellar_1965}. This feature is generally attributed to carbonaceous grains such as polycyclic aromatic hydrocarbons (PAHs) \citep[e.g.,][]{joblin_contribution_1992, bradley_astronomical_2005, shivaei_uv_2022} or nano-sized graphite grains \citep{li_infrared_2001}. Prior to the launch of the \emph{James Webb Space Telescope (JWST)}, spectroscopic detections of the UV bump beyond the local Universe were limited to metal-enriched galaxies at $0.01 \leq z \leq 3$ \citep[e.g.,][]{noll_presence_2007, noll_gmass_2009, shivaei_uv_2022}. Following the launch of \emph{JWST}, PAH emission was detected at $z\sim4$ in a dust-rich galaxy, confirming the presence of carbonaceous dust grains in the early Universe. Subsequently, \emph{JWST} enabled
the surprising detection of a handful of UV bump galaxies up to $z=7.55$ \citep[][]{witstok_carbonaceous_2023, markov_dust_2023, markov_evolution_2025, markov_resolved_2025, fisher_rebels-ifu_2025, ormerod_detection_2025} when the Universe was only $\sim 700$ Myr old.

On average, the UV bump strength is expected to decline with increasing redshift \citep{markov_evolution_2025}, yet individual galaxies with a bump strength $A_{\lambda\rm{,max}}\gtrsim0.4$ mag have been detected at $z\sim7$ \citep[e.g.,][]{witstok_carbonaceous_2023, markov_dust_2023, markov_evolution_2025, ormerod_detection_2025}, indicating rapid carbonaceous dust enrichment within the first billion years of cosmic time. Interestingly, both \citet{witstok_carbonaceous_2023} and \citet{ormerod_detection_2025} report a shift in the peak wavelength of the UV bump to wavelengths longer than typically observed within the MW, which could indicate a differing dust grain composition at high-redshift \citep{blasberger_observational_2017}, such as larger PAH molecules \citep[e.g.,][]{Lin_2023,Lin_2025}.
However, the peak wavelength is not seen to be offset in all UV bump galaxies at high redshift \citep[see Appendix A;][]{ormerod_detection_2025} which may suggest a diversity in UV bump properties at this epoch, although larger sample sizes and better signal-to-noise spectra are needed to determine how significant this variation is. 

The properties of these high redshift UV bump galaxies are derived from rest-frame UV and optical spectra obtained with the \emph{JWST}/Near-infrared Spectrograph \citep[NIRSpec;][]{jakobsen_near-infrared_2022, ferruit_near-infrared_2022, boker_-orbit_2023}, therefore tracing only the unobscured star formation. Consequently, the star formation rate (SFR) estimates for these galaxies may be significantly underestimated \citep[e.g.,][]{madau_cosmic_2014}. At lower redshifts ($z\leq4$), it is well established that the majority of star formation is obscured by dust \citep[][]{zavala_evolution_2021}, and results at higher redshifts indicates that $\sim30\%$  of star formation remains obscured at $z\sim7$ \citep[][]{algera_alma_2023}. Furthermore, individual galaxies at this epoch have been found to be up to $90\%$ obscured in rest-UV observations \citep[e.g.,][]{hygate_alma_2023}, implying that a substantial fraction of star formation may still be undetected in the Epoch of Reionisation (EoR). As a result, our understanding of the properties of UV bump galaxies and their contribution to the cosmic star formation rate density (SFRD), particularly in the early Universe, remains incomplete.

To address the limitations of rest-frame UV and optical observations, which are affected by dust obscuration, far-infrared (FIR) emission lines such as the fine-structure transition of ionised carbon (C$^+$) at 158 {\um}  ({\cii} $158${\um}) provide a valuable dust-unobscured probe of the early Universe. Under typical ISM conditions, {\cii} is the dominant cooling line and is thought to originate primarily from photodissociation regions (PDRs) \citep[][]{hollenbach_photodissociation_1999, wolfire_photodissociation_2022}. PDRs are predominantly neutral regions of the ISM where far-UV photons regulate the heating and chemistry \citep[][]{hollenbach_photodissociation_1999}.
Additionally, {\cii} has been shown to trace the total SFR in `normal' galaxies \citep{de_looze_applicability_2014}, with little evolution seen over the past 13 Gyr of cosmic time \citep{schaerer_alpine-alma_2020}. By combining {\cii} observations with rest-UV observations, it becomes possible to quantify the fraction of obscured star formation and therefore gain a more complete understanding of star formation and galaxy evolution in the early Universe. 

In addition to tracing obscured star formation, FIR continuum observations provide constraints on the dust properties of high-redshift galaxies \citep[e.g.,][]{inami_alma_2022, Schouws_2022, sommovigo_dust_2021, sommovigo_alma_2022, witstok_dual_2022, witstok_empirical_2023, Witstok_25_ALMA, bakx_accurate_2021, Bakx_2026_pixiedust, algera_alma_2023,algera_2026_dust}. While the UV bump is associated with small carbonaceous grains, the FIR continuum traces the emission from larger dust grains. 
Together these observations allow us to explore the physical origin and evolution of dust grain properties, such as the relative grain size distribution, within the early Universe. For example, a weak FIR continuum relative to a strong UV bump feature suggests a relative abundance of small grains compared to large grains, which could indicate efficient reprocessing in the ISM that converts large grains to small dust grains \citep[e.g.,][]{narayanan_ultraviolet_2024}. Conversely, a strong detection of the FIR dust continuum would indicate the presence of large dust grains, which would instead favour stellar sources as the key dust production mechanism \citep{galliano_interstellar_2018}. Despite the detection of the UV bump in several $z>6$ galaxies, a direct link to their FIR dust properties is yet to be established. 

Here, we present Northern Extended Millimeter Array (NOEMA) observations targeting the {\cii} $158${\um} emission line and dust continuum in {\id} at $z=7.1$. This study represents the first direct attempt at linking FIR observations to the UV bump within the Epoch of Reionisation. 
This paper is organised as follows: in Section \ref{sec: observations} we discuss the observations used in this work, in Section \ref{sec: methods} we discuss our method and analysis, and we discuss our findings in Section \ref{sec: discussion}. Finally, our findings are summarised in Section \ref{sec: summary}. Throughout this paper we assume a standard cosmology of $H_0 = 70~{\rm km~s}^{-1}{\rm Mpc}^{-1}$, $\Omega_m = 0.3$, and $\Omega_\Lambda = 0.7$ (where 1\,arcsec corresponds to 5.18\,kpc at the redshift of our target), and a solar abundance of $12+\mathrm{log(O/H)} = 8.69$ \citep{asplund_chemical_2021}. All magnitudes are quoted in the AB magnitude system \citep{oke_secondary_1983}.  We use the \citet{kroupa_2001} initial mass function (IMF) where applicable. 

\section{Observations}
\label{sec: observations}

{\id} was first identified as a Lyman break galaxy (LBG) in \citet{bouwens_uv_2015}, and as a \emph{Spitzer}/IRAC-excess source in \citet{roberts-borsani_z_2016} with an IRAC colour of $[3.6]-[4.5]=0.5^{+1.5}_{-0.5}$. Follow-up Keck/MOSFIRE observations revealed Lyman-$\alpha$ (Ly$\alpha$) emission in {\id} $(\mathrm{EW_{0,Ly\alpha}} = 16\pm8 ~\textrm{\AA}, f(\mathrm{Ly\alpha})=(1.50\pm0.71)\times10^{-18}~ \rm erg~s^{-1}~cm^{-2})$, spectroscopically confirming a redshift of $z_{\mathrm{Ly}\alpha} = 7.10813\pm0.00019$ \citep[][]{Roberts_Borsani_2023_Lya}.
\emph{JWST}/NIRSpec observations were obtained as part of the NIRSpec Wide Guaranteed Time Observations (GTO) Program \citep[Wide;][]{maseda_nirspec_2024}, with NIRCam imaging taken as part of the \emph{JWST} Advanced Deep Extragalactic Survey \citep[JADES;][]{eisenstein_overview_2023} from program 1181 (PI: Eisenstein). For a more detailed description of these observations, we refer the reader to \citet{ormerod_detection_2025}.
Recently, {\id} was identified as a UV bump galaxy in \citet{ormerod_detection_2025}, exhibiting the strongest known $2175${\AA} UV bump absorption feature in its rest-UV spectrum in a $z>4$ galaxy ($A_\mathrm{\lambda,~max}=0.46^{+0.06}_{-0.07}$~mag), with a tentative $\sim3\sigma$ shift in the peak wavelength of the bump ($\lambda_\mathrm{max}=2257^{+26}_{-28}${\AA}).

\subsection{NOEMA Observations}
\label{sec: NOEMA observations}
We obtained 13.1 hours of NOEMA observations as separate programs over two observing cycles, with the first observed in 2024, and the second carried out in 2025-2026. We summarise the observations below, and show the observations from each program in Appendix \ref{sec:noema_appendix}.

\textbf{2024:} 
{\id} was observed for 2.6 hours in configuration 12C-N020+N01 (program S24CM, PIs: A. de Graaff \& K. Ormerod) on 2024 October 31, with precipitable water vapour values of $1-3$ mm. The observations covered  frequency ranges of $232.18-240.31$ GHz (upper side band, USB) and $216.70-224.82$ GHz (lower side band, LSB), corresponding to redshift intervals of $z=6.91-7.19$ and $z=7.45-7.77$, respectively.  Some minor flagging of the data were required due to weather conditions. 

\textbf{2025-26:}
Follow-up observations of {\id} were obtained, being observed for 10.5 hours in configuration 12C (program W25EG, PIs K. Ormerod \& A. de Graaff) on 2025 December 27 and 2026 January 1, with precipitable water vapour values of $<2$ mm and good weather conditions throughout. The observations covered  frequency ranges of $232.18-240.31$ GHz (USB) and $216.70-224.82$ GHz (LSB), corresponding to redshift intervals of $z=6.91-7.19$ and $z=7.45-7.77$, respectively. Very little flagging of the data were required. 

All data reduction and calibration were carried out with the \textsc{gildas} software, with support from IRAM astronomers. Prior to further analysis, we combined both programs in the $uv$-plane. 

We create a {\cii} data cube from the combined USB data, covering the frequency range $233.41-235.36$ GHz. We then taper the data cube using a circular $150$m taper to obtain a more accurate measurement of the total {\cii} flux and clean with 100 iterations prior to imaging the data cube. We followed the cleaning procedures advised by IRAM astronomers, and did not impose a cleaning threshold. 
This results in a median rms in the data cube of $0.46$ mJy in a $38$ km~s$^{-1}$ channel, with a beam size of $1.30\arcsec \times 1.25\arcsec$.  
We merge the USB and LSB before making the continuum image, and exclude the frequency range corresponding to $\pm250~\mathrm{km}~\mathrm{~s}^{-1}$ around the {\cii} emission line in the USB.

\subsection{\emph{JWST} Observations}
\label{sec: JWST observations}
{\id} was observed with NIRSpec as part of the NIRSpec Wide GTO survey \citep[Wide;][]{maseda_nirspec_2024}, and with Near-infrared Camera \citep[NIRCam;][]{rieke_jades_2023} observations taken as part of the \emph{JWST} Advanced Deep Extragalactic Survey \citep[JADES;][]{eisenstein_overview_2023} from programme 1181 (PI: D. Eisenstein). 
The NIRSpec observations -- consisting of PRISM/CLEAR (PRISM hereafter), G235H, and G395H observations --are reduced using the same reduction pipeline as NIRSpec Multi-Object Spectroscopy (MOS) GTO surveys \citep[e.g.,][]{Curtis-Lake_23, cameron_jades_2023, Carniani24, bunker2024, saxena24} developed by the ESA NIRSpec Science Operations Team (SOT) and GTO teams, as described in \citet{Carniani24}. The Wide GTO reduction differs slightly from other GTO reduction pipelines by skipping the sigma-clipping algorithm used to exclude outliers when creating the 1D spectrum. For full details of the Wide data reduction pipeline, we refer the reader to \citet{maseda_nirspec_2024}.
Finally, we apply the same aperture correction to the PRISM spectrum as described in \citet{ormerod_detection_2025}.

Additionally, the NIRSpec PRISM spectrum is available from the DAWN \emph{JWST} Archive \citep[DJA;][]{heintz_strong_2024}, which is reduced using \texttt{msaexp} \citep{brammer_msaexp_2023, de_graaff_rubies_2025}. The DJA version 4 (v4) reduction extends the PRISM wavelength coverage from $5.3\mu\rm m$ to $5.5\mu\rm{m}$ \citep[][]{Pollock_2025_DJAv4, Valentino_2025_DJA_v4}, enabling the detection of the H$\alpha$ emission line in {\id}. Following the method laid out in \citet{ormerod_detection_2025}, we apply an aperture correction to the DJA spectrum, resulting in a good agreement between the GTO and DJA reduction pipelines. 
We present a comparison of the NIRSpec Wide GTO spectrum and the DJA spectrum in Appendix \ref{sec:nirspec_comp}.

\section{Methods and Analysis}
\label{sec: methods}

\subsection{{\cii} 158 $\mu$m Emission Line}
\label{sec: CII emission line}

\begin{figure}
    \centering
    \includegraphics[width=0.95\columnwidth]{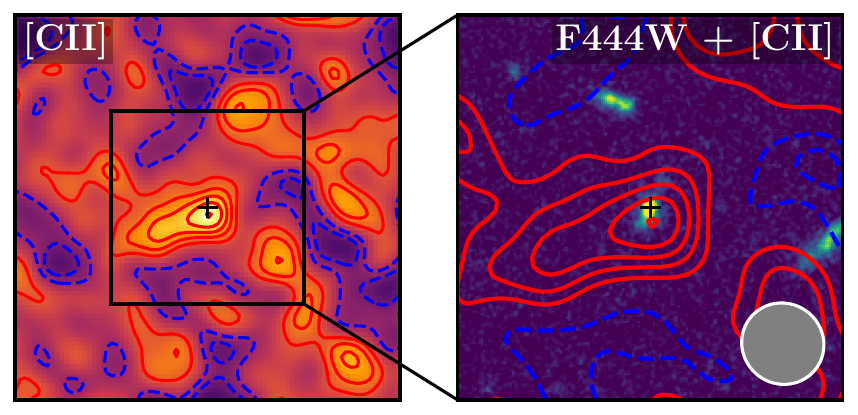}
    \caption{\textit{Left}: $12\arcsec \times 12\arcsec$ panel showing the NOEMA USB data collapsed over the frequency range $234.37 - 234.46$ GHz ($\Delta v = -53$ to $+63~\mathrm{km~s^{-1}}$ relative to the line peak). 
   \textit{Right}: $6\arcsec \times 6\arcsec$ panel showing the \emph{JWST}/NIRCam F444W cutout of {\id} with {\cii} contours overlaid. The grey ellipse in the bottom right corner indicates the beam size.
  In both panels, the solid red contours show the $1\sigma,~2\sigma,~3\sigma$, $4\sigma$, and $5\sigma$ levels of the {\cii} image, while the blue dashed contours show the $-1\sigma$ and $-2\sigma$ levels. The black cross shows the targeted coordinates. North is up and East is to the left.}
    \label{fig:cii cutout}
\end{figure}

\begin{figure}
    \centering
    \includegraphics[width=\columnwidth]{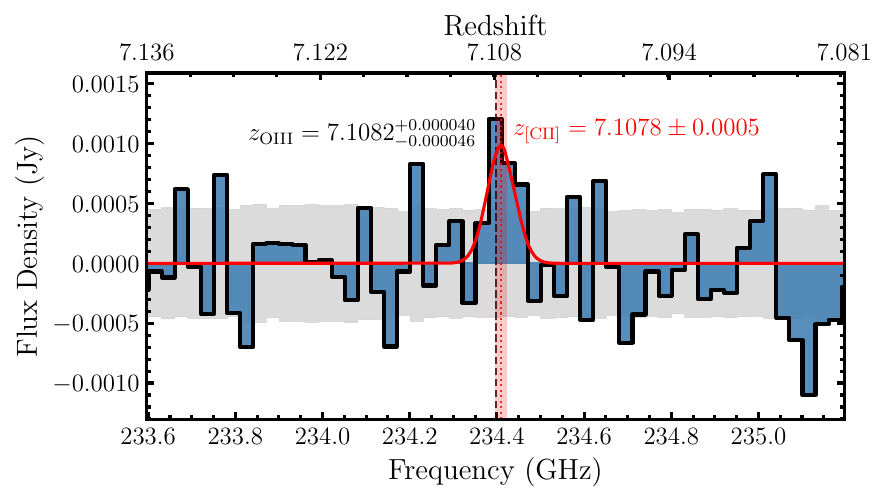}
    \caption{The spectrum of {\id}, obtained by summing the flux within the $2\sigma$ contours of the tapered {\cii} image that overlap with the position of {\id}. We detect {\cii} at $5.1\sigma$ with a redshift of $z_\mathrm{[C\textsc{ii}]} = 7.1078 \pm 0.0005$. The solid red line shows the best-fit Gaussian emission line profile and the grey shading represents the measured RMS of each channel. The black dashed line indicates the redshift determined from \emph{JWST}/NIRSpec G395H observations ($z_\mathrm{[O\textsc{iii}]}$), while the red dotted line indicates the redshift of the {\cii} emission line. The red shading shows the uncertainty on the derived redshift, highlighting the agreement with $z_\mathrm{[O\textsc{iii}]}$.}
    \label{fig:CII spectrum}
\end{figure}

We create the {\cii} image shown in Figure \ref{fig:cii cutout} by collapsing the {\cii} data cube over the frequency range $234.37 - 234.46$ GHz. From this, we find a peak SNR of $5.1$ in excellent spatial coincidence with the location of {\id}. We extract a spectrum by summing the flux of all pixels with $\rm SNR>2$ that overlap the position of the target in the {\cii} image, with the resulting spectrum shown in Figure \ref{fig:CII spectrum}. 
From the best-fit Gaussian line profile, we obtain a redshift of $z_{\mathrm{[C}\textsc{ii}\mathrm{]}} = 7.1078 \pm 0.0005$, as shown in Figure \ref{fig:CII spectrum}, in excellent agreement with $z_\mathrm{[O\textsc{iii}]} = 7.1082^{+0.000040}_{-0.000046}$ obtained from \emph{JWST}/NIRSpec G395H observations.

To further test the robustness of this detection, we apply the Matched Filtering in 3D (\textsc{mf3d}) algorithm, which is designed for blind emission line searches in interferometric data cubes \citep[][]{Pavesi_2018}.  \textsc{mf3d} models emission as a three-dimensional Gaussian distribution in both the spatial and spectral dimensions, and performs a filtering algorithm in Fourier space. 
This yields a detection of the same signal at $\mathrm{SNR}=3.8$. 
We estimate the detection purity by assuming that all negative peaks arise from statistical fluctuations \citep[][]{Pavesi_2018}. In the full data cube, covering $45.9\arcsec \times 45.9\arcsec$ over $1.95$ GHz, we identify $704$ negative features with $|\rm{SNR}|\geq3.8$. However, as in \citet{molyneux_spectroscopic_2022,Witstok_25_ALMA}, this was not a blind search across the full data cube, but restricted to a much smaller volume defined by the known spectroscopic redshift of $z_\mathrm{[O\textsc{iii}]} = 7.1082^{+0.000040}_{-0.000046}$.  Searching within a radius of $0.5\arcsec$ over a velocity range of $\pm100$ km s$^{-1}$, the expected number of features with SNR $\geq3.8$ is $0.023$. Therefore, the probability that the observed feature is due to a statistical fluctuation is $\sim2.3\%$, resulting in an estimated purity of $\sim97.7\%$. The difference in the SNR from \textsc{mf3d} may stem from \textsc{mf3d} modelling emission as three-dimensional Gaussian distributions, whereas the emission in {\id} may deviate from this parametrisation.  

We adopt a Bayesian fitting procedure to model the {\cii} emission line as a Gaussian line profile, using a \textsc{python} implementation of the \textsc{multinest} nested sampling algorithm \citep{feroz_multinest_2009}, \textsc{pymultinest} \citep{buchner_x-ray_2014}. 
We calculate the {\cii} luminosity from the {\cii} emission line flux, using 
\begin{equation}
\label{eqn:cii lum}
L_{\mathrm{[C\textsc{ii}]}}=1.04 \times 10^{-3} \times S_{\mathrm{[C\textsc{ii}]}} \Delta v D_L^2 v_{\text {obs }} L_{\odot},
\end{equation}
where $S_{\mathrm{[C\textsc{ii}]}} \Delta v$ is the line flux in Jy km s$^{-1}$, $D_L$ is the luminosity distance in Mpc, and $v_{\text {obs }}$ is the observed frequency in GHz \citep{solomon_warm_1992, carilli_cool_2013,Smit_Bowler_2026}. The star formation rate is then estimated using the following relation from \citet{de_looze_applicability_2014}: 
\begin{equation}
\label{eqn:delooze14}
\log \mathrm{SFR} / M_{\odot} \mathrm{yr}^{-1}=-6.99 +1.01 \times \log L_{[\mathrm{C\textsc{ii}}]} / L_{\odot} .
\end{equation}
The resulting {\cii} SFR is presented in Table \ref{tab:sfrs} and shown in Figure \ref{fig:SFR_comparison}.

We estimate the dynamical mass as in \citet{Ubler_GANIFS_2023}: 
\begin{equation}
\label{eqn: Mdyn}
M_{\mathrm{dyn}}=K(n) K(q) \frac{\sigma^2 R_e}{G},
\end{equation}
where $K(n)=8.87-0.831 n+0.0241 n^2$ with $n$ as the Sérsic index \citep[][]{Cappellari_2006}, $K(q) = [0.87 + 0.38e^{-3.71(1-q)}]^2$ with $q$ as the axis ratio \citep{van_der_Wel_2022}, giving $K(n) \times K(q)=4.92^{+1.01}_{-1.13}$ for the adopted structural parameters, $R_e$ is the effective radius, $G$ is the gravitational constant, and $\sigma$ is the integrated stellar velocity dispersion ($\sigma_\mathrm{[C\textsc{ii}]} = 42^{+16}_{-15}~\rm km~s^{-1}$). 
We adopt the the structural parameters of the main `Sérsic 1' component
from \citet{ormerod_detection_2025}, as it contains the majority of the galaxy's stellar mass. 
From this, we find a dynamical mass of $\log(M_\mathrm{dyn}/M_\odot) = 8.95^{+0.51}_{-0.66}$, with the large uncertainty arising from the uncertainty in the Sérsic index of {\id}. 
We find that the dynamical mass is slightly lower than that obtained using the velocity dispersion of the [O\textsc{iii}]$\lambda5007$ emission line ($\sigma_\mathrm{[O\textsc{iii}]} = 77\pm3~\rm km~s^{-1}$ from fitting to the G395H spectrum), which using Equation \ref{eqn: Mdyn} gives $\log(M_\mathrm{dyn, O\textsc{iii}}/M_\odot) = 9.47\pm0.17$. The estimate derived from the [O\textsc{iii}] emission line is likely higher than the [C\textsc{ii}] estimate due to non-circular motion \citep[][]{Phillips_2025, kohandel_dynamically_2024}.

The properties of {\id} are presented in Table \ref{tab:properties}.

{\renewcommand{\arraystretch}{1.5}
\begin{table}
    \centering
    \caption{Properties of {\id}. Error bars represent $1\sigma$ uncertainties, and upper limits are 3$\sigma$ upper limits. We provide both $z_{\mathrm{[O\textsc{iii}],5007}}$, measured from the G395H spectrum, and $z_\mathrm{prism}$, measured from the $R\sim100$ PRISM spectrum \citep[][]{ormerod_detection_2025}.}
    \begin{tabular}{l|c} 
         Parameter & Value \\ \hline
         Wide ID & 2008001576\\
         R.A. &12:37:37.941\\
         Dec. &+62:20:22.850\\
         $M_\mathrm{UV}$ & $-20.29\pm0.05$ \\
         $z_{\mathrm{[O\textsc{iii}],5007}}$ & $7.1082^{+0.000040}_{-0.000046}$ \\
         $z_\mathrm{prism}$ & $7.11235$ \\
         $z_{\mathrm{[C\textsc{ii}]}}$ & $7.1078 \pm 0.0005$\\
         $L_{\mathrm{[C\textsc{ii}]}} ~(10^8~L_\odot)$ & $1.32^{+0.46}_{-0.44}$\\
         FWHM$_{\mathrm{[C\textsc{ii}]}}$ (km s$^{-1}$) & $99^{+38}_{-35}$\\
         $\log(\rm M_\mathrm{dyn}/M_\odot)$ & $ 8.95^{+0.51}_{-0.66}$\\
         $\log(\rm M_\star/M_\odot)$ & $8.39^{+0.13}_{-0.09}$ \\
         $158\mu\rm m$ continuum flux ($\mu$Jy) & $<29$\\
         {\cii} flux (Jy km s$^{-1}$) & $0.11\pm0.04$\\
         \hline
         \end{tabular}
    \label{tab:properties}
\end{table}}

\subsection{Dust Continuum}
\label{sec: dust continuum}

\begin{figure}
    \centering
    \includegraphics[width=0.7\columnwidth]{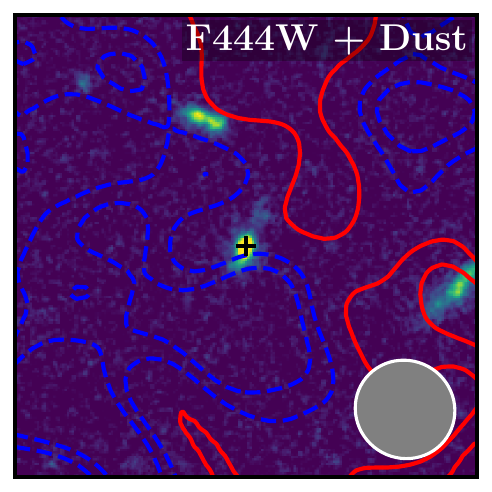}
    \caption{$6\arcsec \times 6\arcsec$ cutout of the F444W image, overlaid with contours derived from the dust continuum image. The blue dashed lines indicate the $-1\sigma$ and $-2\sigma$ levels, while the red solid lines show the $1\sigma$ contours. The ellipse in the bottom right corner represents the beam size for the combined dust continuum image. The black cross shows the target coordinates. North is up and East is to the left.}
    \label{fig:dust cutout}
\end{figure}

We detect no continuum emission at $3\sigma$ within $1\arcsec$ of {\id}, and therefore place a $3\sigma$ upper limit of $<29$ $\mu$Jy on the dust continuum, obtained from $3~\times$ RMS of the continuum image (Figure \ref{fig:dust cutout}). To estimate the corresponding upper limit on the infrared luminosity ($L_\mathrm{IR}$), we consider greybody spectral energy distribution (SED) templates from the Bayesian fitting code \textsc{mercurius} \citep[Multimodal Estimation Routine for the Cosmological Unravelling of Rest-frame Infrared Uniformised Spectra;][]{witstok_dual_2022, witstok_empirical_2023}. We adopt dust temperatures of $T_\mathrm{dust} = 40$K and $T_\mathrm{dust} = 60$K, and emissivity parameters $1.5 \leq \beta_{\mathrm{IR}} \leq 2$, motivated by observational and theoretical studies of galaxies at $z>6$ \citep[e.g.,][]{bakx_accurate_2021, Witstok_2023_Dust, sommovigo_alma_2022}.
All templates assume an optically thin dust opacity model.  

For $\beta_\mathrm{IR}=2$, we obtain upper limits of $L_\mathrm{IR} \leq 3.2\times10^{11}L_\odot$ and $L_\mathrm{IR} \leq 13.4\times10^{11}L_\odot$ for  $T_\mathrm{dust} = 40$K and $T_\mathrm{dust} = 60$K, respectively.
The corresponding infrared star formation rate (SFR$_\mathrm{IR}$) upper limits are estimated adopting a prescription of SFR$_\mathrm{IR}~(M_{\odot} \mathrm{yr}^{-1}) = 1.28\times10^{-10}L_\mathrm{IR}/L_\odot$ as in \citet{inami_alma_2022}, and converted to a Kroupa IMF. The SFR$_\mathrm{IR}$ upper limits are presented in Table \ref{tab:sfrs} and shown in Figure \ref{fig:LCII_SFR}. 
Additionally, we use \textsc{mercurius} to estimate the dust mass. We assume a dust temperature of $T_\mathrm{dust}=50~\rm K$ \citep[e.g.,][]{sommovigo_alma_2022, witstok_empirical_2023} and a dust-emissivity index of $\beta_\mathrm{IR}=2$, which results in a dust mass upper limit of $M_\mathrm{dust} \lesssim 5.3\times10^6 M_\odot$.

\subsection{SED Fitting}
\label{sec:sed fitting}
We model the SED of {\id} by simultaneously fitting both the \emph{JWST} photometry and spectroscopy, employing \textsc{bagpipes} v1.3.2 \citep[Bayesian Analysis of Galaxies for Physical Inference and Parameter EStimation;][]{Carnall_2018, Carnall_2019}, with \citet{bc03} (BC03) stellar models and the \citet{kroupa_2001} IMF.
We largely follow the SED fitting procedure in \citet{ormerod_detection_2025}, with two updates, which are summarised below. 
First, as \textsc{bagpipes} v1.3.2 now enables the fitting of potential damped Lyman-$\alpha$ (DLA) absorption following the prescription in \citet{Witstok_2025_Lya}, we do not mask the Ly-$\alpha$ region during the fitting. We refer the reader to \citet{Witstok_2025_Lya} for full details of the DLA fitting procedure and summarise below.
The absorption cross section of hydrogen is modelled as the Voigt profile approximation from \citet{tasitsiomi_2006}, with a linear correction from \citet{Lee_2013} applied. We fix the redshift of the foreground DLA system to the systemic redshift and assume a uniform prior of $19 < \log_{10}(N_\mathrm{H\textsc{i}}~\rm /~cm^{-2}) < 24$, assuming the temperature of the absorbing gas to be $T=1000$ K. 
Second, the \textsc{bagpipes} nebular models have been updated to \textsc{cloudy} v25.00, with the removal of ISM dust grains resulting in emission lines up to $\sim1$ dex brighter at ionisation parameters $\log_{10}U > -2$ than previously predicted. As a result, we do not mask the [O\textsc{iii}] emission line doublet. However, we have verified that masking the emission lines yields consistent results. 
We utilise the same aperture correction derived in \citet{ormerod_detection_2025}. 

We fix the redshift to that obtained from the PRISM spectrum \citep[$z_\mathrm{prism}=7.11235$;][]{ormerod_detection_2025} rather than the value obtained from the high-resolution spectrum ($z_\mathrm{[O\textsc{iii}],5007}=7.1082^{+0.000040}_{-0.000046}$), as recent studies have reported a small wavelength calibration offset between the NIRSpec PRISM and high-resolution modes \citep[e.g.,][]{de_graaff_rubies_2025, deugenio_jades_2025, scholtz_jades_2026}. We have verified that the results are consistent if we allow the redshift range to vary within a small range centred on $z_\mathrm{prism}$.

We employ a non-parametric star formation history (SFH) from \citet{Leja19}, which fits the SFRs in fixed time bins, with $\Delta$logSFR between bins linked by a Student's t-distribution. We fit a `bursty continuity' model, commonly adopted in studies of high-redshift galaxies \citep[e.g.,][]{tacchella_stellar_2022}, with scale $\sigma = 1$ and $\nu = 2$ degrees of freedom. The first time bins are fixed to $0~\mathrm{Myr} < t < 3~\rm{Myr}$ and $3~\mathrm{Myr} < t < 10~\rm{Myr}$. 
These bins are chosen to enable the SFH to capture extreme line emission and to allow for a bursty SFH, motivated by observational evidence at high-redshift \citep[e.g.,][]{boyett_extreme_2024, endsley_burstiness_2024, looser_jades_2025}, while also allowing for smooth or declining SFHs if favoured \citep[e.g.,][]{harvey_epochs_2025}. The remaining four bins are equally log-spaced in lookback time until $t(z=20)$. The inferred SFH is largely insensitive to the number of bins used as long as $N_\mathrm{bins}\geq4$ \citep{Leja19}. 

We modify \textsc{bagpipes} to reduce the default SFR timescale to $10~$Myr to account for the increased specific star formation rate (sSFR), compared to galaxies at lower redshift, and also measure SFRs averaged over timescales of 3 and 30 Myr.
The allowed total stellar mass formed and stellar metallicity is allowed to vary uniformly between $10^5~\rm{M}_\odot < M_\star < 10^{15} ~\rm{M}_\odot$, and $0<Z_*<2.0~ \mathrm{Z}_{\odot}$, respectively. Nebular emission is included using a grid of \textsc{cloudy} \citep{ferland_2017_2017} models, parametrised by the ionisation parameter $(-3 < \rm{log}_{10} U < -0.5)$, which \textsc{bagpipes} computes self-consistently. 

We adopt the \citet{salim_dust_2018} dust attenuation curve, which parametrises the dust curve shape with a power-law deviation $\delta$ from the \citet{calzetti_dust_2000} model ($\delta = 0$ for the Calzetti curve, with negative values of $\delta$ producing steeper slopes) and includes a Drude profile to model the 2175\AA\ bump. We allow the bump strength to vary uniformly between $0 < B < 10$, assuming a fixed central wavelength of 2175{\AA} rest-frame and bump width of 250{\AA}. 
We use a Gaussian prior on the V-band dust attenuation $(A_V)$ with $\mu_{A_V} = 0.15$ mag, $\sigma_{A_V} = 0.15$ mag and attenuation limited to $0 < A_V < 7$ mag, fixing the fraction of attenuation arising from stellar birth clouds to $60\%$, with the remaining $40\%$ coming from the diffuse ISM \citep{chevallard_simulating_2019}. We also assume a uniform prior on the velocity dispersion in the range $1-1000$ km~s$^{-1}$. Finally, we assume that the spectrum follows the PRISM resolution curve based on a point-source morphology, using the resolution curve of an idealised point source generated with \texttt{msafit}, as described in \citet[][Appendix A]{de_graaff_ionised_2024}.

We summarise the priors used in Table \ref{tab:priors} and present the inferred properties of {\id} in Table \ref{tab:bagpipes_results}. The best-fit spectrum and inferred SFH are shown in Figure \ref{fig:bagpipes_sed} and Figure \ref{fig:bagpipes_sfh}. We revise the emission line fluxes, ratios, and inferred gas-phase metallicity presented in \citet{ormerod_detection_2025}, with the updated $A_V$ value derived in this work, as described in Appendix \ref{sec: emission line appendix}.

\subsection{Additional Star Formation Rate Tracers}
\label{sec:sfr tracers}

To complement the  SFR$_\mathrm{IR}$ and SFR$_\mathrm{[C\textsc{ii}]}$ estimates derived in this work, we incorporate further SFR estimates to probe a range of star formation timescales. 
We first include SFR$_{10}$ and SFR$_{30}$ from the SED fitting carried out in Section \ref{sec:sed fitting}, tracing the star formation averaged over 10 and 30 Myr, respectively.
We also include SFR$_\mathrm{UV}$ (uncorrected for dust) as a tracer of the unobscured star formation over the last $\sim100$ Myr, which we estimate using the \citet{madau_cosmic_2014} relation, converted to a Kroupa IMF. 

Furthermore, the extended spectral coverage of the DJA v4 spectrum enables us to measure the H$\alpha$ emission line in {\id} for the first time, allowing us to estimate both the H$\alpha$ SFR (SFR$_{{\mathrm{H}\alpha}}$) and the Balmer decrement. 
We carry out emission line fitting of the H$\alpha$ and H$\beta$ emission lines following the method in \citet{ormerod_detection_2025}. Briefly, we create Gaussian models on an oversampled wavelength grid and convolve them with the LSF of an idealised point source from \citet{de_graaff_ionised_2024}.
As the H$\alpha$ emission line is blended with the [N\textsc{ii}]$\lambda\lambda6548, 6584$~{\AA} doublet due to the PRISM resolution, we therefore fit a triple Gaussian \citep[e.g.,][]{rowland_rebels-ifu_2026}. To reduce the number of free parameters in the fit, we tie the line width of all three lines and fix the central wavelength of the emission lines. Finally, we fix the flux ratio of the [N\textsc{ii}] doublet to $3.049$ \citep{Storey_2000}. 
From this, we find H$\alpha$/H$\beta = 2.88\pm0.28$.

We then correct the H$\alpha$ flux for dust attenuation in two ways.
Firstly, we apply a dust correction based on the $E(B-V)_\mathrm{gas}$, calculated using the Balmer decrement: 
\begin{equation}
    E(B-V)_\mathrm{gas} = \frac{2.5}{k(\lambda_\mathrm{H\beta}) - k(\lambda_\mathrm{H\alpha})} \log \left (\frac{(\mathrm{H}\alpha/\mathrm{H}\beta)_\mathrm{obs}}{2.86} \right)
\end{equation}
where $k(\lambda)$ is the dust attenuation curve, which is related to the attenuation magnitude ($A(\lambda)$) and colour excess ($E(B-V)$) by $k(\lambda)=A(\lambda)/E(B-V)$. Finally, $(\mathrm{H}\alpha/\mathrm{H}\beta)_\mathrm{obs}$ is the observed flux ratio. We assume an intrinsic H$\alpha$/H$\beta$ flux ratio of 2.86, assuming Case B recombination with an electron temperature of $T_e=10^4$~K and an electron density of $n_e=10^2~\mathrm{cm}^{-3}$ \citep[][]{Osterbrock_1989, Osterbrock_2006}. 
From the measured Balmer decrement, we obtain $E(B-V)=0.007_{-0.007}^{+0.093}$, resulting in $A_V = 0.021_{-0.021}^{+0.288}$. 

Secondly, we apply a dust correction assuming the value of $A_V$ obtained from SED fitting in Section \ref{sec:sed fitting}. We convert the stellar $A_V$ to a nebular $A_V$ following \citet{reddy_mosdef_2020}. In both cases, we correct for dust attenuation assuming the \citet{cardelli_relationship_1989} attenuation curve. We then estimate the H$\alpha$ SFR following \citet{kennicutt_star_2012} as a tracer of star formation over $\sim10$~Myr.

\begin{figure}
    \centering
    \includegraphics[width=0.95\columnwidth]{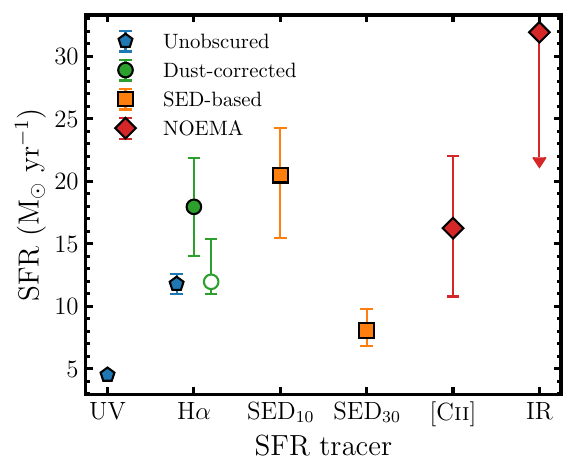}
    \caption{Comparison of star formation rates for {\id} obtained from various SFR tracers. We show unobscured SFRs (not corrected for dust attenuation) as blue pentagons, dust corrected SFRs as green circles, SED-based SFRs as orange squares, and measures of SFR obtained with NOEMA as red diamonds. The open green circle represents the H$\alpha$ SFR obtained from the dust correction based on the Balmer decrement, while the solid green circle represents the H$\alpha$ SFR obtained from the SED-based dust correction. 
    The SFR$_\mathrm{IR}$ upper limit shown assumes a dust temperature of $T_\mathrm{dust}=40$K.}
    \label{fig:SFR_comparison}
\end{figure}

We find SFR$_\mathrm{H\alpha}$ to be in good agreement with SFR$_{10}$ when assuming the value of $A_V$ from SED fitting, despite the SED fitting of the Wide spectrum not including coverage of the H$\alpha$ emission line. The agreement between  SFR$_\mathrm{H\alpha}$ and SFR$_\mathrm{10}$ is also seen in the REBELS-IFU survey \citep{Fisher_SF_2026} at $z=6.5-7.7$.
Star formation rates are presented in Table \ref{tab:sfrs} and shown in Figure \ref{fig:SFR_comparison}. 

{\renewcommand{\arraystretch}{1.5}
\begin{table}
    \centering
    \caption{Star formation rates of \id\ from different tracers. $\rm SFR_{UV}$ is not corrected for dust attenuation.
    H$\alpha$ (SED) and H$\alpha$ (Balmer dec.) are corrected for dust attenuation using $A_V$ from SED fitting (converted to nebular $A_V$) and the Balmer decrement, respectively. We also present the $\rm H\alpha$ SFR prior to dust correction.
    SED$_{10}$ and SED$_{30}$ are obtained from \textsc{bagpipes} SED fitting and averaged over 10 and 30 Myrs, respectively. Uncertainties are reported at the $1\sigma$ level.}
    \begin{tabular}{l|c} 
         Tracer & SFR ($\rm M_\odot$ yr$^{-1}$) \\ \hline
         UV & $4.52^{+0.21}_{-0.20}$\\ 
         H$\alpha$ (SED)& $18.0\pm3.9$\\ 
         H$\alpha$ (Balmer dec.)& $12.0^{+3.5}_{-1.0}$\\
         H$\alpha$ (uncorrected) & $11.8\pm0.8$\\
         SED$_{10}$ & $20.5^{+3.8}_{-5.0}$\\
         SED$_{30}$ & $8.1^{+1.7}_{-1.3}$\\
         {\cii} & $16.2^{+5.8}_{-5.5}$\\
         IR ($T_\mathrm{dust}=40$K) & $\leq 32$\\
         IR ($T_\mathrm{dust}=60$K) & $\leq 134$\\
         \hline
         \end{tabular}
    \label{tab:sfrs}
\end{table}}

\section{Discussion}
\label{sec: discussion}

\subsection{Star Formation Rate Tracers}
\label{sec:obscured SF}

\begin{figure*}
    \centering
    \includegraphics[width=0.8\textwidth]{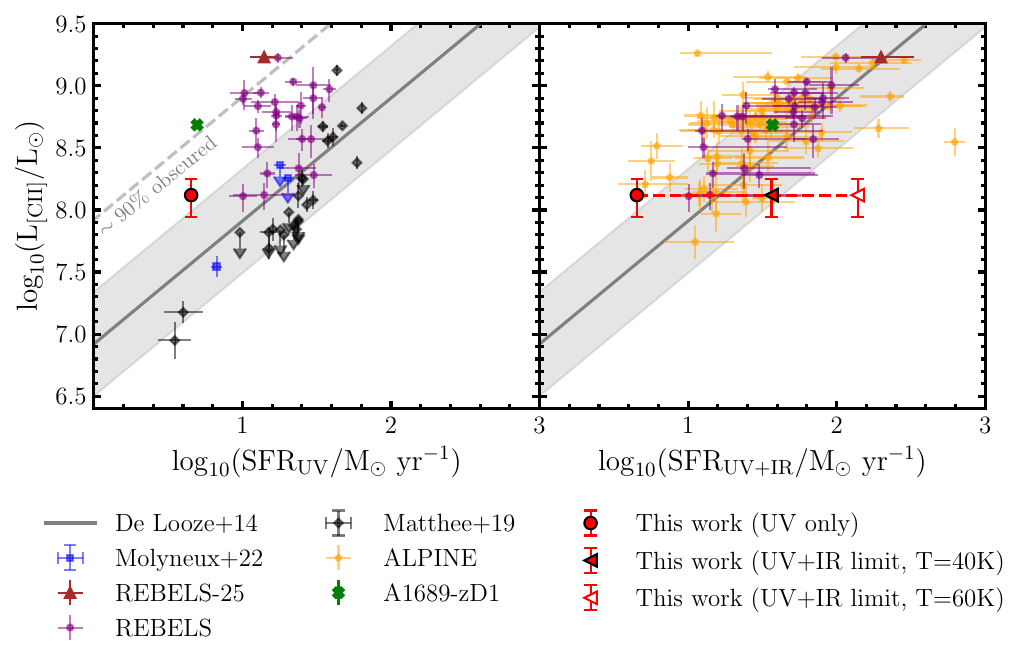}
    \caption{\textit{Left:} {\cii} luminosity as a function of SFR$_{\mathrm{UV}}$. \textit{Right:} {\cii} luminosity as a function of SFR$_{\mathrm{UV+IR}}$. The SFR$_\mathrm{UV}$ of {\id} is shown as a red circle, with $1\sigma$ error bars. The upper limits of the UV+IR SFRs are also shown in the right panel, for $T_\mathrm{dust}=40$K and $T_\mathrm{dust}=60$K. The red dashed line shows the possible range for the the SFR. For comparison, we include individual galaxies from \citet{molyneux_spectroscopic_2022}, the REBELS survey (Schouws et al, in prep.), the ALPINE survey \citep{bethermin_alpine-alma_2020}, and a compilation of $z\sim6-7$ sources from \citet{matthee_resolved_2019}. All limits shown are $3\sigma$ upper limits. The $1\sigma$ upper limits from \citet{matthee_resolved_2019} are converted to $3\sigma$ upper limits for consistency. We also show the galaxies REBELS-25 \citep[][]{hygate_alma_2023, rowland_rebels-25_2024} and A1689-zD1 \citep[][]{watson_dusty_2015, bakx_accurate_2021}. The $L_{\mathrm{[C\textsc{ii}]}}$-SFR relation from \citet{de_looze_applicability_2014} is indicated by the solid grey line, with the errors shown by the grey shaded region. The grey dashed line in the left panel indicates where galaxies with $\sim90\%$ obscured star formation would lie.} 
    \label{fig:LCII_SFR}
\end{figure*}

In Figure \ref{fig:LCII_SFR}, we present the measured SFR$_\mathrm{UV}$ (uncorrected for dust) and SFR$_\mathrm{UV+IR}$ (a measure of total SFR) of {\id}, alongside a comparison to sources and relations from the literature.  A tight $L_\mathrm{[C\textsc{ii}]}$-SFR relationship is seen in the local Universe, as shown by \citet{de_looze_applicability_2014} (see Equation \ref{eqn:delooze14} and Figure \ref{fig:LCII_SFR}). At higher redshifts, a good agreement has been found with the local \citet{de_looze_applicability_2014} relation (henceforth referred to as the `local relation') when considering  SFR$_\mathrm{UV+IR}$ \citep[e.g.,][]{schaerer_alpine-alma_2020, van_leeuwen_characterising_2024}. However, when the IR contribution to the overall SFR is not accounted for, sources with significant dust obscuration can deviate considerably from the local relation. For example, both REBELS-25 and A1689-zD1, which host significant obscured star formation \citep[$>87\%$][]{Akins_2022, hygate_alma_2023}, show offsets to the local relation of $>1$ dex. 
We find that {\id} lies $0.55$ dex above the local relation. Although this suggests the presence of obscured star formation, the fraction of obscured star formation is not as significant as highly obscured galaxies such as REBELS-25 or A1689-zD1, consistent with the non-detection of the dust continuum in {\id}. When adopting a dust temperature of $T_\mathrm{dust}=40\mathrm{K}$ and accounting for the $\rm SFR_{IR}$ contribution, {\id} lies within the scatter of the local relation. 

Alternatively, the slight excess {\cii} luminosity seen in the left panel of Figure \ref{fig:LCII_SFR}, may reflect enhanced photoelectric heating boosting the luminosity of the {\cii} emission line, rather than obscured star formation. At depths of $A_V \lesssim 5$, photoelectric ejection of electrons from small grains and PAHs is the dominant gas heating process within PDRs \citep[][]{bakes_photoelectric_1994}, with half of the heating coming from small grains ($<15${\AA}).
Cooling of atomic gas mainly occurs through collisional excitation and radiative de-excitation of atomic fine-structure lines, predominantly through the {\cii} fine-structure line \citep{wolfire_photodissociation_2022}. As such, the increased heating within the PDR due to an abundance of PAHs could enhance the {\cii} emission within {\id} and contribute to the offset from the local relation. 

We compare the SFRs obtained from various tracers in Figure \ref{fig:SFR_comparison}.  We find SFR$_\mathrm{UV} = 4.52^{+0.21}_{-0.20}~M_{\odot} \mathrm{yr}^{-1}$ (not corrected for dust), and use this to quantify the obscured star formation. In order to quantify the fraction of obscured star formation we define the fraction of obscured star formation as the 
difference between the SFR inferred from the {\cii} and UV emission, assuming {SFR}$_{\mathrm{[C\textsc{ii}]}}$ = SFR$_\mathrm{UV+IR}$, over the SFR from {\cii}: 
\begin{equation}
    f_{\mathrm{obs}} =
\frac{\mathrm{SFR}_{\mathrm{[C\textsc{ii}]}} - \mathrm{SFR}_{\mathrm{UV}}}
     {\mathrm{SFR}_{\mathrm{[C\textsc{ii}]}}}.
\end{equation}
We define $f_\mathrm{obs}$ in this way due to the non-detection of the dust continuum, which only allows an estimate of an upper limit of SFR$_\mathrm{IR}$. 
From this, we find the fraction of obscured star formation to be $f_\mathrm{obs} \sim 0.72$. We note that this is consistent with using the upper limit of SFR$_\mathrm{IR}$ assuming $T_\mathrm{dust}=40$~K, which results in $f_\mathrm{obs} < 0.86$.  This is consistent with other galaxies at $z\sim7$, such as the REBELS-IFU sample with obscured fractions of $f_\mathrm{obs}=0.56-0.78$ \citep[][]{Fisher_SF_2026}. However, it is important to note that the obscured fraction may decrease if {\cii} emission is enhanced due to an abundance of small dust grains or if a significant fraction of {\cii} emission is associated with the ionised gas phase of the ISM (see below). 

\begin{figure}
    \centering
    \includegraphics[width=0.9\columnwidth]{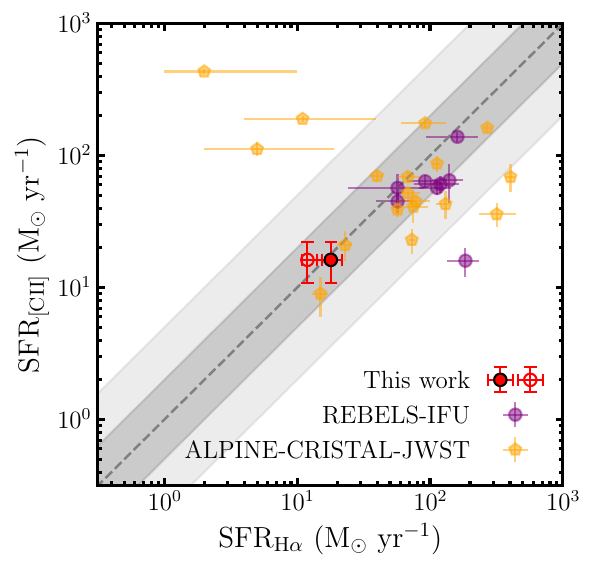}
    \caption{$\rm SFR_{H\alpha}$ vs $\rm SFR_{[C\textsc{ii}]}$. We show $\rm SFR_{H\alpha}$ as a solid red marker.
    with the open marker indicating $\rm SFR_{H\alpha,BD}$. 
    We compare to the REBELS-IFU sample \citep[purple circles;][]{Fisher_SF_2026} and the ALPINE-CRISTAL-JWST survey \citep[orange pentagons;][]{Faisst_2026}. The grey dashed line shows the one-to-one relation, while the dark (light) grey shading shows a factor of 2 (5) from the one-to-one relation.}
    \label{fig:sfr_cii_ha}
\end{figure}

We find that {\sfrcii} is consistent with SFR$_{10}$ and SFR$_\mathrm{H\alpha}$ within $1\sigma$ (see Figure \ref{fig:SFR_comparison} and Figure \ref{fig:sfr_cii_ha}). Similarly, \citet{Faisst_2026} find a good agreement between SFR$_\mathrm{H\alpha}$ and {\sfrcii} to first order, with increasing scatter at higher H$\alpha$ SFR attributed to bursty star formation histories. While this agreement between tracers is expected for sources with a constant star formation history, the inferred SFH for GNWY-7379420231 shows strong fluctuation within the first three bins of lookback time, and therefore a lower {\sfrcii} compared to SFR$_{10}$ and SFR$_\mathrm{H\alpha}$ might have been expected. Given that {\cii} emission is thought to be dominated by the PDR, where photons between 6 and 13.6 eV dominate the physical properties \citep{wolfire_photodissociation_2022},  {\sfrcii} is most likely to be linked with star-formation on slightly longer timescales, e.g. SFR$_{30}$. Figure \ref{fig:SFR_comparison} shows an SFR$_{30}$ of $8.1^{+1.7}_{-1.3}~M_\odot~\rm yr^{-1}$, just below {\sfrcii}, though still consistent within 2$\sigma$. If this difference is real, a possible explanation could be that most of the {\cii} emission is emerging from ionised regions, or perhaps the enhanced photo-electric heating of the PDR due to small dust grains is boosting the {\cii} emission. 

\subsection{Physical Properties}
\label{sec: physical properties}

The {\cii} luminosity can be used to estimate the total gas mass, $M_\mathrm{gas}$, of a galaxy using the {\cii} conversion factor, $\alpha_{[\text {C } \textsc{ii}]}$, although it has also been suggested that {\cii} may trace H\textsc{i} gas \citep[e.g.,][]{heintz_measuring_2021,heintz_alma_2022}.
Furthermore, the reliance on the adopted conversion factor introduces systemic uncertainties due to the wide range of reported values \citep{vallini_spatially_2025}, ranging from $\alpha_{[\text {C } \textsc{ii}]}\approx 72 ~M_{\odot} / L_{\odot}$ for local metal poor galaxies \citep{madden_tracing_2020} to $\alpha_{[\text {C } \textsc{ii}]} < 10 ~M_{\odot} / L_{\odot}$ in $z\sim4.5$ lensed dusty star forming galaxies (DSFGs) \citep{rizzo_dynamical_2021},  $z\sim6-7$ quasars \citep{kaasinen_cold_2024} and $z>4$ galaxies in \citet{sommovigo_dust_2021}. 

To infer the gas mass for {\id}, we use the conversion factor $\alpha_{[\text {C } \textsc{ii}]}=7_{-1}^{+4} M_{\odot} / L_{\odot}$ derived for DSFGs from \citet{rizzo_dynamical_2021}. \citet{vallini_spatially_2025} find an anti-correlation between $\alpha_{[\text {C } \textsc{ii}]}$ and metallicity, suggesting $\alpha_{[\text {C } \textsc{ii}]} \sim 6.8M_{\odot} / L_{\odot}$ for {\id}, supporting the adopted value.
We then estimate a gas mass of $\log _{10}(M_\mathrm{gas}/M_\odot) = 8.97^{+0.33}_{-0.24}$, in agreement with the inferred dynamical mass within $1\sigma$. 
We note that adopting a scaling relation with a higher $\alpha_\mathrm{[C\textsc{ii}]}$ value would yield a gas mass that exceeds the estimated dynamical mass derived from the [O{\sc iii}] and {\cii} line widths (section \ref{sec: CII emission line}). 

While the inferred gas mass is high, it remains consistent with the dynamical mass within $1\sigma$. This suggests that {\id} is a gas-rich system (stellar mass fraction of 28\%), with the large gas mass providing fuel for ongoing star formation.

\subsection{Dust Production in the Early Universe}
\label{sec: dust discussion}

The 2175{\AA} UV bump is a broad absorption feature seen in the rest-UV spectrum of some galaxies, often attributed to carbonaceous dust grains such as PAHs. As such, the identification of the UV bump in galaxies up to $z=7.55$ provides important constraints on dust properties and evolution in the early Universe \citep[][]{witstok_carbonaceous_2023, markov_dust_2023, markov_evolution_2025, markov_resolved_2025, fisher_rebels-ifu_2025, ormerod_detection_2025}, and can be used to probe dust grain properties within the first billion years of cosmic time. For example, the UV bump of {\id} shows a central wavelength of $\lambda_{\max }=2257_{-28}^{+26} ${\AA}, $2.9\sigma$ higher than the peak wavelength seen in the MW curve. This is similar to the peak wavelength of $\lambda_{\max }=2263_{-24}^{+20}${\AA} measured for JADES-GS-z6-0 in \citet{witstok_carbonaceous_2023}, and may indicate that the dust grains responsible for the UV bump feature have a larger molecular size than that in the MW \citep[e.g.,][]{blasberger_observational_2017, Li_2024_UV_Bump, Lin_2025}. 

Additionally, the presence of the UV bump within the first billion years of cosmic time is thought to challenge existing models of dust formation. SED fitting in Section \ref{sec:sed fitting} suggests the presence of a very young stellar population inferred from fitting to the integrated spectrum and photometry ($t_\star\sim14$ Myr). If a stellar population of this age is considered in isolation, it would imply that AGB stars are unlikely to be responsible for carbonaceous dust production in {\id}, due to the $\sim 300$ Myr timescale required to evolve off the main sequence \citep[e.g.,][]{schneider_formation_2024}. 
For consistency with the integrated SED fitting in this work, we repeat the SED fitting of the three morphological components carried out in \citet{ormerod_detection_2025} with BC03 stellar templates, and find substantial stellar mass build up at ages $>300$~Myr within the most massive `Sérsic 1' component ($\log _{10}\left(\mathrm{M}_* / \mathrm{M}_{\odot}\right)=7.86_{-0.65}^{+0.17}$), consistent with that found in \citet{ormerod_detection_2025}. The updated stellar masses of each component result in a stellar mass ratio of $6:1$ between the main `Sérsic 1' component and the two merging components, consistent with the interpretation of {\id} as a minor merger.
Alternatively, recent theoretical work by \citet{Matsumoto_2026} finds that the UV bump becomes more prominent on a timescale of $\sim250$~Myr through the formation of small grains produced through the interplay between shattering and accretion. 

If the presence of an older stellar population is confirmed, this could indicate that the required carbonaceous dust grains may already be present, and are illuminated by the recent burst of merger-induced star formation. If this is the case, the existing NIRSpec slit placement may capture the region with the strongest UV bump feature in {\id}, explaining the strength of the UV bump without requiring anomalous dust production mechanisms or properties. Upcoming \emph{JWST} Cycle 5 NIRSpec IFU observations (PI: K. Ormerod; PID: 11433) will enable the underlying stellar populations and the localised nature of the UV bump feature to be better constrained. These resolved observations will provide a direct test of whether the UV bump is a galaxy-wide feature, or a localised property visible in the existing spectrum due to slit placement. The increased SNR of the integrated spectrum will also improve constraints on the width and offset of the peak wavelength of the UV bump, shedding light on the properties of the dust grains responsible for this feature at early times.

While the FIR continuum traces large dust grains, the UV bump is attributed to small grains, such as PAHs or nano-sized graphite grains \citep[e.g.,][]{joblin_contribution_1992, li_infrared_2001, bradley_astronomical_2005, shivaei_uv_2022},  which may form in a top-down formation pathway via shattering of larger grains (formed in SNe) through grain-grain collisions \citep{Jones_1996}.
The non-detection of the FIR dust continuum in {\id}, despite the strong UV bump, favours a scenario in which the dust grain size distribution is biased towards small carbonaceous dust grains. 
This may suggest that efficient reprocessing in the ISM converts large dust grains into the small grains responsible for the UV bump feature \citep[e.g.,][]{narayanan_ultraviolet_2024, narayanan_pah_lifecycle_2026} through top-down PAH formation in the turbulent, diffuse ISM where large grains are shattered into small grains. This alters the grain size distribution within {\id}, while the dust mass remains the same. However, it should be noted that the lack of a dust continuum detection does not suggest the complete absence of all large grains, only that the grain size distribution is biased towards small carbonaceous grains. 
The proposed \emph{PRobe far-Infrared Mission for Astrophysics (PRIMA)} telescope will enable observations of PAH emission in the EoR for the first time, providing a direct test of their presence at early times  \citep[e.g.,][]{yoon_polycyclic_2025}.

Alternative pathways for the production of the dust responsible for the UV bump at high redshift have also been suggested. Recent analysis of the UV bump identified in \citet{witstok_carbonaceous_2023} suggests that the PAHs are unlikely to originate solely from AGB stars or shattering of large dust grains in the ISM \citep{Nanni_2025}, indicating instead that additional PAH formation in the ISM or contributions from massive stars may be necessary.
Furthermore, \citealt{Nanni_2025} find that a large PAH fraction (4\%-4.6\%) is required to reproduce the UV bump, suggesting that substantial PAH mass is able to survive in high-redshift galaxies. 

From the $3\sigma$ dust mass upper limit ($M_\mathrm{dust} < 5.3\times10^6 M_\odot$), we find a dust-to-stellar mass ratio of $M_\mathrm{d}/M_* < 2\%$, or $\log_{10}(M_\mathrm{d}/M_*) \sim -1.67$. 
A dust-to-stellar mass ratio of $\sim2\%$ would require rapid dust enrichment,
with the dust mass likely growing through accretion of gas-phase metals onto dust grains in the ISM once a critical metallicity has been reached, which may occur at around $10-20\%$ solar metallicity \citep[e.g.,][]{asano_dust_2013, remy-ruyer_gas--dust_2014, remy-ruyer_linking_2015, roman-duval_metal_2022}. Recent work by \citet{algera_2026_dust} suggests that ISM dust growth may be an important process in the build up of dust reservoirs in EoR galaxies.
However, the true dust mass and dust-to-stellar mass ratios are likely much lower, with typical dust-to-stellar mass ratios at $z\sim7$ of $<1\%$ \citep[e.g.,][]{Algera_2024a}. Alternatively, the true stellar mass of {\id} may be slightly underestimated due to the effect of outshining in integrated SED fitting, with a total stellar mass of $\log_{10}(\rm M_\star/M_\odot)=8.65^{+0.21}_{-0.23}$ obtained from updating the resolved SED fitting in \citet{ormerod_detection_2025} with BC03 stellar templates.
Moreover, the dust mass upper limit is derived from a single band observation and may underestimate the true dust mass, highlighting the need for further observations to better constrain the dust SED.

\section{Summary}
\label{sec: summary}
We have presented the NOEMA detection of the {\cii} $158$\um\ emission line in the UV bump galaxy {\id}, in a first attempt to link FIR properties to a $z>6$ UV bump detection. Although we do not detect the FIR dust continuum, we are able to place constraints on the dust properties of {\id}, and present the first detailed multi-wavelength analysis of a UV bump host galaxy at $z>7$.
Our main findings are summarised as follows: 

\begin{itemize}
    \item We detect {\cii} at $z_\mathrm{[C\textsc{ii}]} = 7.1078 \pm 0.0005$ in {\id}, in excellent agreement with $z_\mathrm{[O\textsc{iii}]} = 7.1082^{+0.000040}_{-0.000046}$ at a significance of $5.1\sigma$. This is supported by the detection of the same signal with the \textsc{mf3d} algorithm with a purity of $97.7\%$. 
    \item We measure a {\cii} luminosity of $L_\mathrm{[C\textsc{ii}]} = 1.32^{+0.46}_{-0.44}\times10^8 ~L_\odot$ in {\id}, which lies $0.55$ dex above the local \citet{de_looze_applicability_2014} relation when comparing to the unobscured SFR. This could suggest the presence of obscured star formation and/or enhanced [C\textsc{ii}] due to an abundance of PAH that efficiently heat the PDR. Furthermore, the [C\textsc{ii}]  based SFR is consistent with the H$\alpha$ SFR and the SED-based SFR averaged over a 10~Myr timescale, which could indicate a significant contribution of [C\textsc{ii}] from ionised gas.
    \item The dust continuum is not detected, therefore we place an upper limit on the obscured star formation of $<32M_\odot\mathrm{yr}^{-1}$, assuming a dust temperature of $T_\mathrm{dust}=40$K. This corresponds to an obscured star formation fraction of $<86\%$. When assuming a total star formation of SFR$_\mathrm{tot}$ = $\mathrm{SFR}_\mathrm{[C\textsc{ii}]}$, we find $f_\mathrm{obs}\sim0.72$.  However, {\sfrcii} may be enhanced by efficient PAH heating, which would suggest a lower true value of $f_\mathrm{obs}$. 
    \item The dust continuum non-detection combined with the detection of a strong UV bump could favour a scenario in which small carbonaceous dust grains, such as PAHs or graphitic grains, are abundant in {\id}.  This supports a top-down formation pathway for PAHs, in which the abundance of large grains is depleted through shattering in the ISM, converting large grains into small grains. Alternatively, the strong bump signature may arise predominantly due to line of sight effects. 
    \item The [C\textsc{ii}] derived dynamical mass of $\log_{10}(\rm M_{\text{dyn }}/M_{\odot})=8.95_{-0.66}^{+0.51}$ and stellar mass of $\log _{10}\left(\mathrm{M}_{\star} / \mathrm{M}_{\odot}\right)=8.39_{-0.09}^{+0.13}$ suggest {\id} is a `normal', moderately massive galaxy with surprisingly mature dust, rather than a heavily dust obscured or massive galaxy.

\end{itemize}

These results highlight the synergy between rest-UV and FIR observations in constraining the dust and gas properties of galaxies in the Epoch of Reionisation. Upcoming JWST Cycle 5 observations will allow a resolved spectroscopic analysis of {\id}, enabling a more in-depth study of dust grain formation mechanisms in a UV bump galaxy in the EoR.

\section*{Acknowledgements}

This work is based on observations carried out under project numbers S24CM and W25EG with the IRAM NOEMA Interferometer. IRAM is supported by INSU/CNRS (France), MPG (Germany) and IGN (Spain). The authors would like to thank IRAM astronomers for their help during the reduction process.
This work is based on observations made with the NASA/ESA/CSA James Webb Space Telescope (JWST). The data were obtained from the Mikulski Archive for Space Telescopes at the Space Telescope Science Institute, which is operated by the Association of Universities for Research in Astronomy, Inc., under NASA contract NAS 5-03127 for JWST. These observations are associated with programmes 1181 and 1211. This study made use of Prospero high-performance computing facility at Liverpool John Moores University. 
Some of the data products presented herein were retrieved from the Dawn \emph{JWST} Archive (DJA). DJA is an initiative of the Cosmic Dawn Center (DAWN), which is funded by the Danish National Research Foundation under grant DNRF140.
KO would like to thank the Science and Technology Facilities Council (STFC) and Faculty of Health, Innovation, Technology and Science (HITS) at Liverpool John Moores University (LJMU) for their studentship. 
RS acknowledges support from a STFC Ernest Rutherford Fellowship (ST/S004831/1). 
JW gratefully acknowledges support from the Cosmic Dawn Center through the DAWN Fellowship. The Cosmic Dawn Center (DAWN) is funded by the Danish National Research Foundation under grant No. 140.
MVM is supported by the National Science Foundation via AAG grant 2205519.
AdG acknowledges support from a Clay Fellowship awarded by the Smithsonian Astrophysical Observatory. AdG acknowledges funding by the European Union (ERC, FIRST-GIANTS, 101221926). AdG is supported by the Lise Meitner Excellence program of the Max Planck Society.
AJB acknowledges funding from the “FirstGalaxies” Advanced Grant from the European Research Council (ERC) under the European Union’s Horizon 2020 research and innovation program (Grant agreement No. 789056).
G.C.J. acknowledges support by the Science and Technology Facilities Council (STFC), ERC Advanced Grant 695671 "QUENCH".

\section*{Data Availability}

The data underlying this article will be shared on reasonable request to the corresponding author.
This work is based on observations made with the NASA/ESA/CSA James Webb Space Telescope (JWST). The data were obtained from the Mikulski Archive for Space Telescopes at the Space Telescope Science Institute, which is operated by the Association of Universities for Research in Astronomy, Inc., under NASA contract NAS 5-03127 for JWST. These observations are associated with programmes 1181 and 1211. Some of the data products presented herein were retrieved from the Dawn \emph{JWST} Archive (DJA). DJA is an initiative of the Cosmic Dawn Center (DAWN), which is funded by the Danish National Research Foundation under grant DNRF140.



\bibliographystyle{mnras}
\bibliography{paper} 




\appendix

\section{NOEMA Observations}
\label{sec:noema_appendix}
\begin{figure}
    \centering
    \includegraphics[width=0.9\columnwidth]{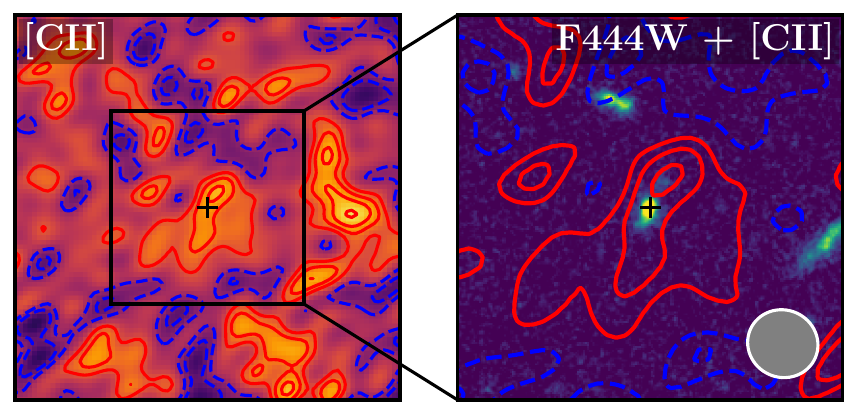}
    \caption{The same as Figure \ref{fig:cii cutout}, but for the 2024 observations (program S24CM) only. }
    \label{fig:CII_2024_Obs}
\end{figure}

\begin{figure}
    \centering
    \includegraphics[width=0.9\columnwidth]{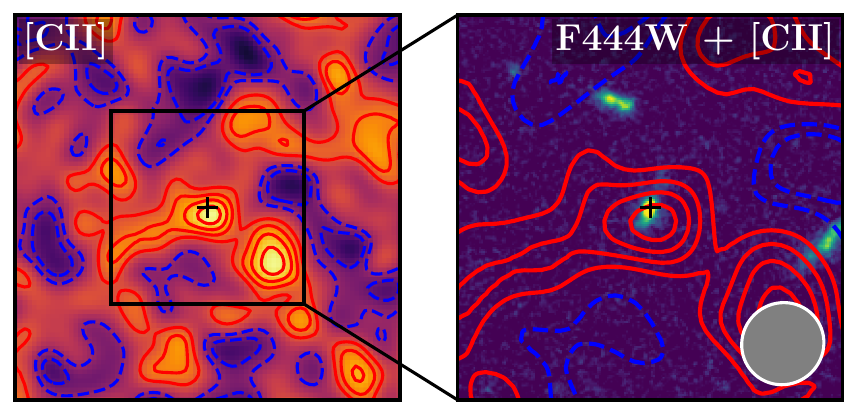}
    \caption{The same as Figure \ref{fig:cii cutout}, but for the 2025-2026 observations (program W25EG) only. }
    \label{fig:CII_2025_26_Obs}
\end{figure}

\begin{figure}
    \centering
    \includegraphics[width=0.98\columnwidth]{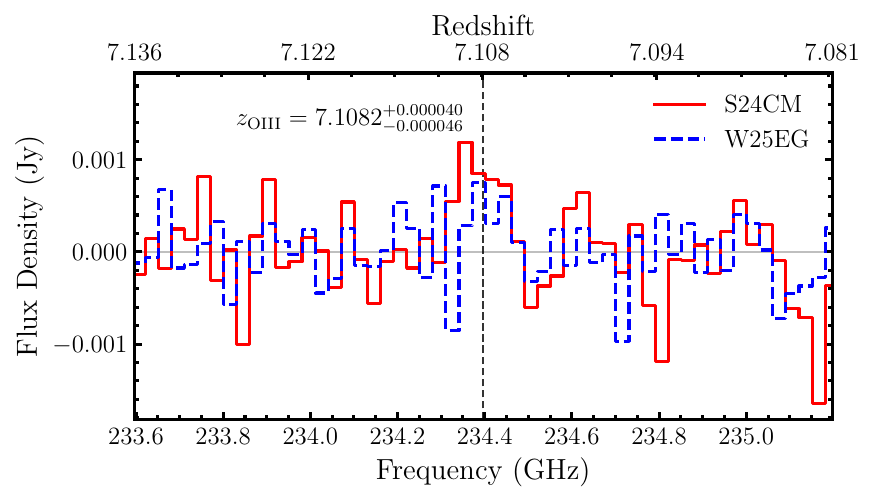}
    \caption{The spectrum of {\id}, obtained by summing the flux within the $2\sigma$ contours of the individual programs. Program S24CM is shown by the solid red line, while program W25EG is shown by the dashed blue line. The black dashed line indicates the redshift determined from \emph{JWST}/NIRSpec G395H observations.}
    \label{fig:CII_spec_appendix}
\end{figure}

We obtained NOEMA observations of {\id} over two observing cycles, with program S24CM observed in 2024, and W25EG observed in 2025-2026. A tentative {\cii} signal $(3.3\sigma)$ is seen in the {\cii} image obtained from the 2024 observations alone, with an on source time of just 2.6 hours. From this, follow-up observations were obtained in program W25EG with 10.5 hours on source, observed over December 2025 to January 2026. We show the {\cii} image from the individual observations in Figure \ref{fig:CII_2024_Obs} and Figure \ref{fig:CII_2025_26_Obs}. The spectra extracted from the $2\sigma$ contours in the {\cii} images from the individual programs are shown in Figure \ref{fig:CII_spec_appendix}.

\section{NIRSpec Comparison}
\label{sec:nirspec_comp}

\begin{figure*}
    \centering
    \includegraphics[width=0.85\textwidth]{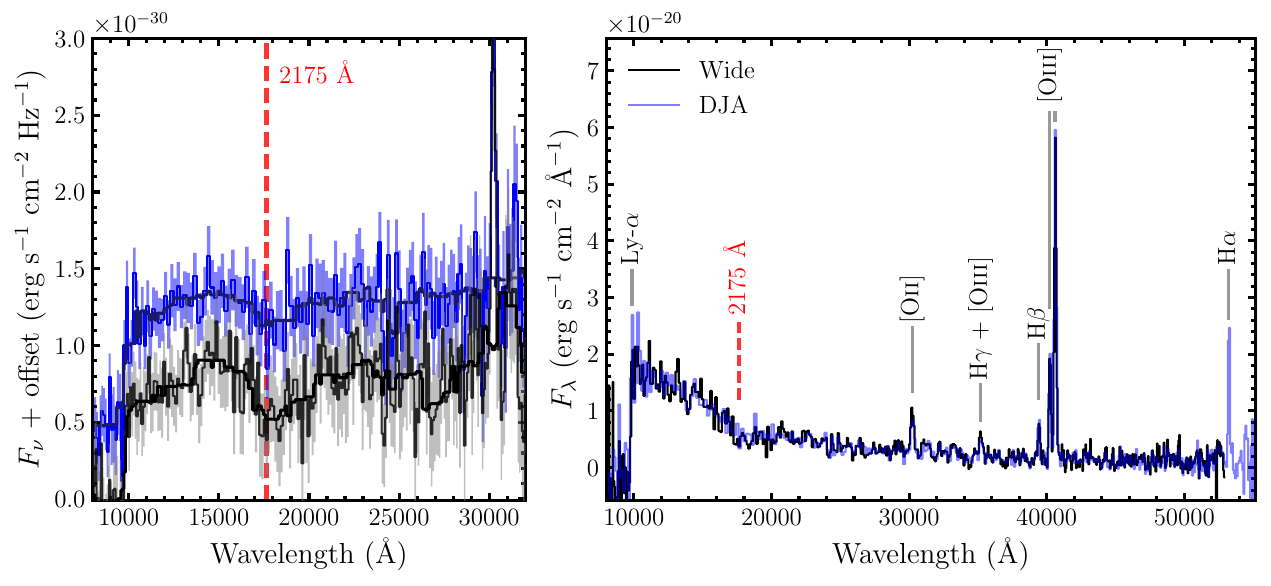}
    \caption{A comparison of the NIRSpec PRISM spectra of {\id} obtained from the NIRSpec Wide GTO reduction pipeline (black) and the DJA reduction pipeline (blue). Shading represents $1\sigma$ uncertainties. \textit{Left:} Zoom in on the UV bump region, with the spectra offset for clarity. The location of the $2175${\AA} UV bump is indicated by the red dashed line. A running median is overlaid over both spectra. \textit{Right:} A comparison of the full spectra. The location of the UV bump is marked with a red dashed line. The location of key emission lines are indicated by grey lines, including the H$\alpha$ emission line, highlighting the extended wavelength coverage offered by the DJA version 4 spectrum. }
    \label{fig:nirspec_comp}
\end{figure*}

We present a comparison of the NIRSpec PRISM spectra of {\id} obtained through differing reduction pipelines in Figure \ref{fig:nirspec_comp}. Both spectra are aperture corrected according to the method from \citet{ormerod_detection_2025}, resulting in a good agreement between reduction pipelines. 

\section{SED Fitting}
\subsection{SED Fitting Priors}
\label{sec: SED fitting priors}
The priors for our \textsc{bagpipes} SED fitting are presented in Table \ref{tab:priors}.

\begin{table}
    \centering
    \caption{Summary of SED fitting parameters and priors used in our \textsc{bagpipes} fitting in Section \ref{sec:sed fitting}. The minimum and maximum values allowed are given in brackets.}
    \begin{tabular}{l|l}
         Parameter & Prior  \\ \hline
         $z$ & Fixed to $z_\mathrm{prism}=7.11235$\\ 
         $\log_{10}(\rm M_\star/M_\odot)$ & Uniform: (5, 15)\\
         $Z_\star/Z_\odot$ & Uniform: (0, 2.0)\\
         $\log_{10}(U)$ & Uniform: (-3, -0.5)  \\ 
         $A_V$ (mag) & Gaussian: (0, 7), $\mu = 0.15, \sigma=0.15$ \\
         $\sigma_v$ (km s$^{-1}$) & Uniform: (1, 1000)\\ 
         $\Delta\log$(SFR)$_i$ & Student's-t: (-50, 50) \\ 
         $B$& Uniform: (0, 10)\\ 
         $\delta$ & Uniform: (-0.5, 0.2) \\ 
         $\log _{10}\left(N_{\mathrm{H\textsc{i}}}\right)$ & Uniform: (19, 24) \\
         
         \hline
    \end{tabular}
    \label{tab:priors}
\end{table}

\subsection{SED Fitting Results}
\label{sec: SED fitting results}
\begin{figure}
    \centering
    \includegraphics[width=0.95\columnwidth]{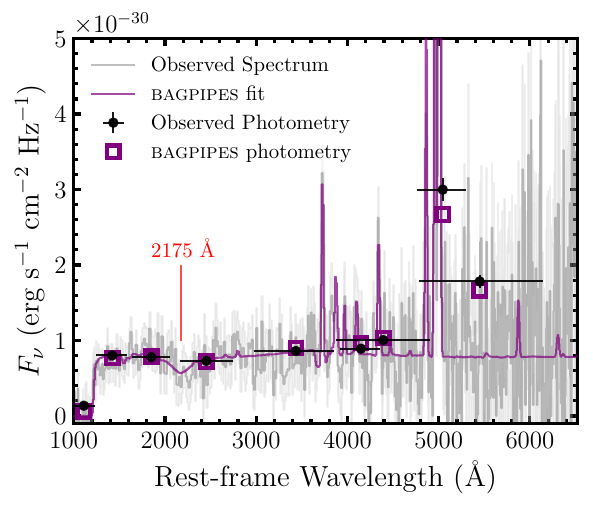}
    \caption{Posterior spectrum obtained from \textsc{bagpipes} SED fitting of the photometry and spectroscopy. The observed spectrum and associated errors are shown in grey, with the observed photometry shown as black circles. The x error bars represent the width of the filter at $50\%$ of the maximum transmission. The posterior spectrum is shown in purple, with the posterior photometry shown as open purple squares. The UV bump feature is clearly seen in both the observed and posterior spectra, and is indicated by a solid red line..}
    \label{fig:bagpipes_sed}
\end{figure}

\begin{figure}
    \centering
    \includegraphics[width=1.0\columnwidth]{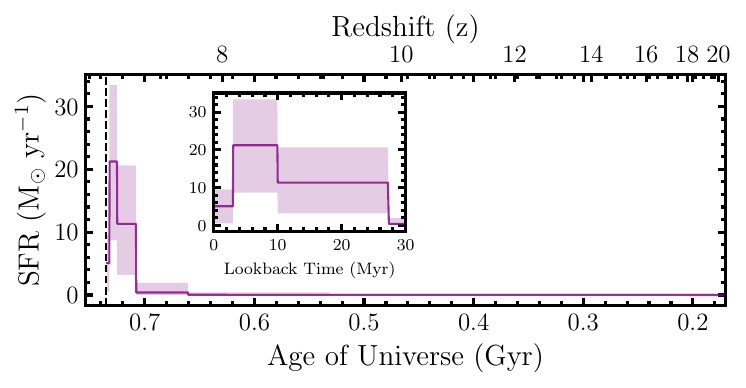}
    \caption{The star formation rate as a function of the age of the Universe is shown by the purple line, with shading representing $1\sigma$ errors. The inset panel shows SFR against lookback time for a zoom in on the three most recent star formation bins. }
    \label{fig:bagpipes_sfh}
\end{figure}

The \textsc{bagpipes} SED fitting results are presented in Table \ref{tab:bagpipes_results}, with the best fit spectrum and photometry shown in Figure \ref{fig:bagpipes_sed}. 
We show the star formation history of {\id} in Figure \ref{fig:bagpipes_sfh}.

{\renewcommand{\arraystretch}{1.5}
\begin{table}
    \centering
    \caption{SED fitting properties of {\id}. The rows are: (1) stellar mass, (2) SFR averaged over 3 Myr, (3) SFR averaged over 10 Myr, (4) SFR averaged over 30 Myr, (5) $V$-band dust attenuation, (6) deviation from the slope of the Calzetti dust curve, (7) UV bump strength in the Salim dust law, (7) stellar metallicity, (8) ionisation parameter, and (9) mass-weighted stellar age. Errors represent $1\sigma$ uncertainties.}
    \begin{tabular}{l|c} 
        Parameter & Value  \\ \hline
        $\log _{10}\left(\mathrm{M}_{\star} / \mathrm{M}_{\odot}\right)$ & $8.39^{+0.13}_{-0.09}$ \\
        SFR$_3$ (M$_\odot$ yr$^{-1}$) & $9.92^{+3.27}_{-1.78}$ \\
        SFR$_{10}$ (M$_\odot$ yr$^{-1}$) & $20.48^{+3.81}_{-5.00}$ \\
        SFR$_{30}$ (M$_\odot$ yr$^{-1}$) & $8.08^{+1.67}_{-1.29}$ \\
        $A_V$ (mag) & $0.27^{+0.10}_{-0.09}$ \\
        $\delta$ & $-0.04^{+0.12}_{-0.18}$ \\
        $B$ & $4.06^{+1.57}_{-1.29}$ \\
        $Z_\star$ ($Z_\odot$) & $0.39^{+0.04}_{-0.05}$ \\
        $\log _{10}(U)$ & $-1.51^{+0.16}_{-0.13}$ \\
        $t_\star$ (Myr) & $14.37^{+39.70}_{-7.98}$ \\ \hline
    \end{tabular}
    \label{tab:bagpipes_results}
\end{table}}

\section{Emission line properties}
\label{sec: emission line appendix}

We apply an updated dust correction to the emission line fluxes from \citet{ormerod_detection_2025}. We apply a dust correction assuming the value of $A_V$ obtained from SED fitting in Section \ref{sec:sed fitting}, converting the stellar $A_V$ to a nebular $A_V$ following \citet{reddy_mosdef_2020}. We then correct for dust attenuation assuming the \citet{cardelli_relationship_1989} attenuation curve. From the NIRSpec PRISM dust corrected emission line fluxes, we derive a range of emission line ratios. Both the emission line fluxes and ratios are reported in Table \ref{tab:line fluxes}. 

We use the emission line ratios to determine the gas-phase metallicity ($Z_\mathrm{neb}$), using the calibrations from \citet{sanders_direct_2024}. We find a value of $Z_\mathrm{neb} = 0.26^{+0.13}_{-0.09}~Z_\odot$. We note that using the \citet{curti_jades_2024} calibrations yields a consistent result ($Z_\mathrm{neb}=0.22^{+0.10}_{-0.09}~Z_\odot$).

\begin{table*}
    \centering
    \caption{\textit{Top:} Dust corrected emission line fluxes measured from the PRISM and G395H spectra. Fluxes are given in units of $10^{-18}$erg s$^{-1}$ cm$^{-2}$. The dust correction is based on the $A_V$ obtained from SED fitting. Emission line fluxes measured from the DJA spectrum rather than the Wide spectrum are indicated.
    [O\,{\sc ii}]$\lambda \lambda 3727\mathrm{,}29$ is blended in the PRISM spectrum, and H$\beta$ is located within the chip-gap in the G395H spectrum. \textit{Bottom:} Emission line ratios obtained from dust corrected NIRSpec PRISM emission line fluxes, measured from the Wide spectrum.}
    \centering
    \begin{tabular}{lcc}
         \textit{Emission line fluxes} & & \\ \hline
         Emission Line & PRISM & G395H \\ \hline
         [O\,{\sc ii}]$\lambda \lambda 3727\mathrm{,}29$& $4.52 \pm 1.97$& -  \\
         $[\mathrm{O}\,\textsc{ii}]\lambda 3727$ & - & $2.51 \pm 1.37$ \\
         $[\mathrm{O}\,\textsc{ii}]\lambda 3729$ & - & $1.84 \pm 1.15$ \\
         H$\beta$ & $2.23 \pm 0.68$ & - \\
         H$\beta$ (DJA) & $2.35 \pm 0.66$ & - \\
         $[\mathrm{O}\,\textsc{iii}]\lambda 4959$ & $6.00 \pm 1.36$ & $6.28 \pm 1.72$ \\
         $[\mathrm{O}\,\textsc{iii}]\lambda 5007$ & $17.7 \pm 3.97$ & $17.7 \pm 3.71$ \\ 
         H$\alpha$ (DJA) & $5.66 \pm 1.24$ & - \\
         \hline
         \textit{Emission line ratios} & & \\ \hline
         Line ratio & Definition & Observed ratio \\ \hline
         R2 & $\log \left([\mathrm{O\textsc{ii}}] \lambda \lambda 3727,3729/\mathrm{H} \beta\right)$ & $0.31\pm0.23$\\
         O32 & $\log \left({[\mathrm{O\textsc{iii}}] \lambda 5007}/{[\mathrm{O\textsc{ii}}] \lambda \lambda 3727,3729}\right)$ & $0.59\pm0.21$\\
         R3 & $\log \left({[\mathrm{O\textsc{iii}}] \lambda 5007}/{\mathrm{H} \beta}\right)$ & $0.90\pm0.16$\\
         R23 & $\log \left({[\mathrm{O\textsc{ii}}] \lambda \lambda 3727,3729+[\mathrm{O\textsc{iii}}] \lambda \lambda 4959,5007}/{\mathrm{H} \beta}\right)$ & $1.10\pm0.18$\\
         $\hat{R}$ & $0.47 \times \mathrm{R}2+0.88 \times \mathrm{R}3$ & $0.94\pm0.18$\\
         \hline
    \end{tabular}
    \label{tab:line fluxes}
\end{table*}


\bsp	
\label{lastpage}
\end{document}